\documentclass[twocolumn,aps]{revtex4}
\input{epsfig.sty}
\newcommand{\beq}{\begin{eqnarray}}
\newcommand{\eeq}{\end{eqnarray}}
\begin{document}
\title{Large Mixing Angle Sterile Neutrinos and Pulsar Velocities }
\author{Leonard S. Kisslinger\\
Department of Physics, Carnegie Mellon University, Pittsburgh, PA 15213\\ 
Ernest M. Henley\\
Department of Physics, University of Washington, Seattle, WA 98195\\
Mikkel B. Johnson \\
Los Alamos National Laboratory, Los Alamos, NM 87545 \\}

\begin{abstract} We investigate the momentum given to a protoneutron star,
the pulsar kick, during the first 10 seconds after temperature equilibrium
is reached. Using a model with two sterile neutrinos obtained by
fits to the MiniBoone and LSND experiments, which is consistent with a new
global fit,  there is a large mixing angle, and the effective volume for 
emission is calculated. Using formulations with neutrinos created by URCA 
processes in a strong magnetic field, so the lowest Landau level has a 
sizable probability, we find that with known parameters the asymmetric 
sterile neutrino emissivity might account for large pulsar kicks.

\end{abstract}
\maketitle
\noindent
PACS Indices:97.60.Bw,97.60.Gb,97.60.JD
\vspace{1mm}

\section{Introduction}

   The existence of sterile neutrinos, in addition to three active neutrinos
of the standard model, is of great interest for both particle physics and 
astrophysics. The Los Alamos LSND experiment found evidence for antineutrino
oscillation\cite{lsnd}. An analysis of LSND, along with other
short-baseline experiments, found\cite{scs04} that the standard model is not
consistent with the data and the best model has two sterile neutrinos in 
addition to three active neutrinos. 

The recent experiment by the MiniBooNE Collaboration\cite{mini} found that 
the data for electron neutrino appearance showed an excess at low energies, 
in comparison to what was expected in the standard model. This data, along 
with the LSND data, has been analyzed in a model with two light sterile 
neutrinos\cite{ms07},  and compared to MiniBooNE data\cite{s07}.  The
mixing angles of two light sterile neutrinos were extracted. (See, however,
Ref.\cite{mm08}, which questions the accuracy of the results of 
Ref\cite{ms07}). An analysis of MiniBooNE, LSND, and other 
experiments\cite{gsm07}, based in part on earlier work with the seesaw 
mechanism\cite{sgmg05}, also found that the preferred model to fit the
data consists of two light sterile neutrinos, in addition to three active
neutrinos.

  Very recently a global analysis of updated MiniBooNE data, along with
data from LSND, KARMEN, NOMAN, Bugey, CHOOZE, CCFR84, and CDHS has been 
carried out\cite{kdcss09}. There are two possible scenerios, both compatible 
with the results of Refs.\cite{ms07,s07}. Fits to appearance and disappearance
with a 3+1 hypothesis have very good $\chi^2$ probabilities, and also
find CPT violation. The CPT violation, with different mixing angles and
mass differences for neutrinos vs antineutrinos, might be understood as
matter effects, similar to the MSW effect\cite{w78,ms85}, but this has not 
yet been proved. Also a 3+2 model was shown to give a good $\chi^2$ 
probability and is compatible within the range of the parameters for the 3+2
model used in Refs.\cite{ms07,s07}, but without the uncertainty pointed
out in Ref.\cite{mm08}.

  In the present work we apply this range of fits\cite{ms07,s07} to 
MiniBooNE/LSND to the study 
of pulsar velocities. The gravitational collapse of a massive star often 
leads to the formation of a neutron star, a pulsar. It has been observed 
that many pulsars move with linear velocities of 1000 km/s or greater,
called pulsar kicks. See  Ref.\cite{hp97} for a review. 
Due to the high density, the active neutrinos have a small mean free path,
and only escape at the surface of the neutrinosphere. During the first ten 
seconds after temperature equilibrium is attained during the gravitational 
collapse of a massive star, the main cooling is via emmision of neutrinos 
produced by the URCA process, and during the next ten seconds via neutrinos 
produced by the modified URCA process. We have investigated the pulsar kicks 
which arise from the modified URCA processes in the time interval 10-20 sec 
after the supernova collapses, when the neutrinosphere is just inside the 
protoneutron star. With a strong magnetic field and temperature, so that the 
population of the lowest Landau level is approximately 0.4 of the total 
occupation probability, and we find large pulsar kicks\cite{hjk07}.

   The largest neutrino emission after the supernova collapse takes place
during the first 10 seconds, with URCA processes dominant. The possibility
of pulsar kicks from anisotropic neutrino emission due to strong magnetic
fields during this time was discussed more than two decades ago\cite{chu84}.
It has been shown\cite{dor} that, with the strength of the magnetic field 
expected during this period, the lowest Landau level has a sizable 
occupation probability, which produces the neutrino emission asymmetry that 
is needed for pulsar kicks. However, due to the high opacity for standard 
model neutrinos in the dense region within the neutrinosphere, few neutrinos 
are emitted, and the large pulsar kick is not obtained\cite{lq98}. 

   Sterile neutrinos with a small mixing angle have small opacities.  
It has been shown\cite{fkmp03} that, using the model of Ref.\cite{dor} and 
assuming the existence of a heavy sterile neutrino (mass $>$ 1 kev) with 
a very small mixing angle constrained to fit dark matter, the pulsar 
kicks could be explained. More recently the effects of such sterile 
neutrinos with large masses and small mixing angles have been studied for 
other processes\cite{k06}. See Ref.\cite{kus09} for a review of processes
that moght be associated with dark matter sterile neutrinos.

 In the present paper we use the fits of Refs.\cite{ms07,s07} with
two sterile neutrinos to investigate the possibility of obtaining
the large pulsar velocities which have been observed. The values of the
masses and mixing angles are within the range found in calculations based
on the seesaw mechanism\cite{gsm07}. Note that
our model differs from that of Ref\cite{fkmp03} in that with a much larger
mixing angle there is a higher probability of sterile neutrinos, but
a much smaller effective volume, due to a larger opacity. However, as
we shall show, since the mean free path is much larger than those of
standard neutrinos, under the conditions in which standard neutrinos 
produce a pulsar velocity of 2-300 km/s, the MiniBoone/LSND sterile
neutrinos can give a kick of more than 1000 km/s.

\section{Asymmetric Sterile Neutrino Emissivity and Pulsar Kicks in Light
Two-Sterile Neutrino Model}

  Within about 1 second after the gravitational collapse of a large star,
the neutrinosphere is formed with a radius of about 40 km, with temperature 
equilibrium. Within about 10 seconds about 98\% of neutrino emission occurs, 
with neutrinos produced mainly by URCA processes. Due to the strong magnetic 
field, neutrino momentum asymmetry is produced within the neutrinosphere, but 
with a small mean free path they are emitted only from a small surface 
layer of the neutrinosphere, and the pulsar kick cannot be accounted for.
If a standard active neutrino, say the electron neutrino, oscillates into
a sterile neutrino, it will escape from the protoneutrino star and 
neutrinosphere, unless it oscillates back into the active neutrino. The
mixing angle plays a key role. In the work of Fuller et al\cite{fkmp03}
the mixing angle is so small that the sterile neutrinos are emitted. In the
present work the starting point is the analysis of MiniBooNE and LSND data,
with the two or more sterile neutrinos with small masses and large mixing 
angles. Before we can proceed, however, it is essential to determine 
possible effects of the high density and temperature of the medium on the 
mixing angles.

\subsection{Mixing Angle in Neutrinosphere Matter}

   It has long been known that dense matter can affect neutrino states.
The famous MSW effect\cite{w78,ms85} for understanding solar neutrinos,
and the study of oscillations of high energy neutrinos\cite{ams05} are
studies of mixing of active neutrinos in matter. There have been many
other studies. In the present work we are dealing with sterile/active 
neutrino mixing  given by the mixing angle $\theta_m$ in neutrinosphere 
matter

\beq
\label{1}
      |\nu_1> &=& cos\theta_m |\nu_e> -sin\theta_m |\nu_s> \\
      |\nu_2> &=& sin\theta_m |\nu_e> +cos\theta_m |\nu_s> \; . 
 \nonumber 
\eeq

  In the work Ref\cite{fkmp03} it was shown that the mixing angle for sterile
neutrinos that can account for dark matter as well as those produced in the 
neutron star core is almost the same as the vacuum value. Starting from the 
much larger mixing angles for the sterile neutrinos that seem to account for
the MiniBoonE, LSND data, we need the value of the mixing angles in the
neutrinosphere, as we discuss below. The effective mixing angle in matter, 
$\theta_m$ can be related to the vacuum mixing angle, $\theta$ by\cite{afp01}
\beq
\label{2} 
          sin^2(2\theta_m) &=& \frac{sin^2(2\theta)}{sin^2(2\theta) +
(cos(2\theta)-\frac{2p V^T}{(\delta m)^2})^2} .
\eeq
In Eq(\ref{2}) $V^T$ is the finite temperature potential, while the
finite density potential due to asymmetries in weakly interacting particles
has been dropped as it vanishes when temperature equilibrium is 
reached\cite{afp01}. A convenient form for $V^T$, with the background of
both neutrinos and electrons included, is given in Ref\cite{nr88}
\beq
\label{3}
        V^T &=& \frac{28 \pi G_F^2}{45 \alpha} sin^2 \theta_W
(1+0.5 cos^2\theta_W) p T^4 \; ,
\eeq
with $G_F,\theta_W$ the standard weak interaction parameters and
$\alpha=1/137$. Assuming T=20$a$ MeV, with $a \leq$1.0, p=$b$ MeV, 
and $(\delta m)^2=$ 1.0 ev$^2$\cite{ms07,mm08}, we find
\beq
\label{4}
 \frac{2p V^T}{(\delta m)^2}& \simeq & 5.1 \times 10^{-3} b^2 a^4 \ll 
cos(2 \theta) \; .
\eeq

Therefore, the mixing angle in the neutrinosphere medium is approximately 
the same as the vacuum mixing angle. This agrees with Ref\cite{fkmp03}.

\subsection{Emissivity With a Light Sterile Neutrino}

   We now use the fits to MiniBooNe and LSND with light sterile neutrinos
to estimate pulsar kicks. The MiniBooNE results are consistent with the 
LSND results and CPT only if there are at least two sterile neutrinos. Models 
with three sterile neutrinos have also been considered\cite{sgmg05,mm08}.
Fits to the MiniBooNE experiment and the LSND results by Ref.\cite{ms07}
in Ref.\cite{s07} with two sterile neutrinos are shown in Fig. 1.
\vspace{-.5 cm}

\begin{figure}[ht]
\begin{center}
\epsfig{file=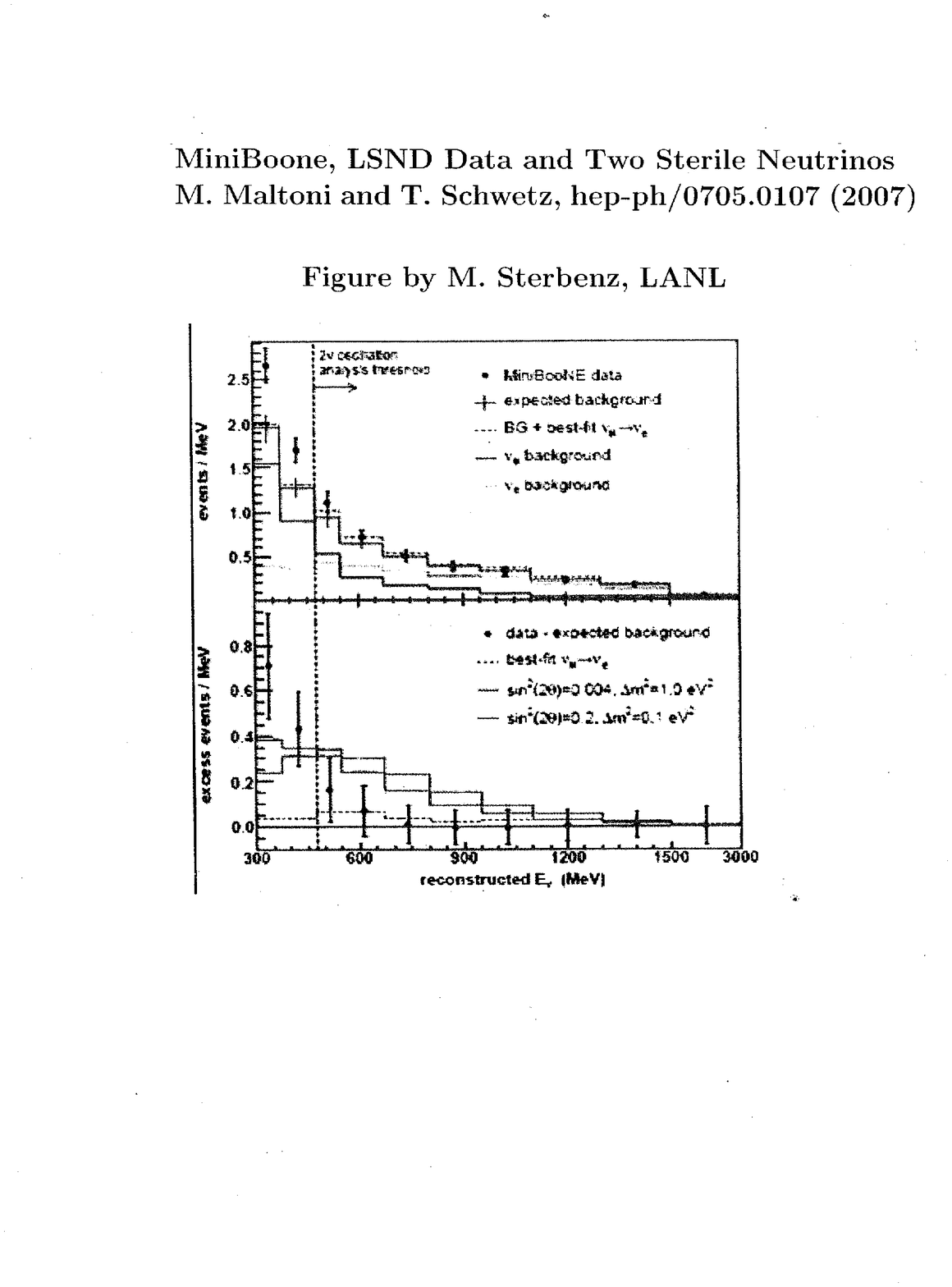,height=11.5cm,width=8cm}
\caption{$sin^2(2\theta_{1s})$=0.004; $sin^2(2\theta_{2s})$=0.2}
{\label{Fig.1}}
\end{center}
\end{figure}
\vspace{3mm}

From the Sterbenz/Maltoni-Schwetz fits one finds for the mixing angles of
the two sterile neutrinos:
\beq
\label{5}
                  (sin 2\theta_{1s})^2 &=& 0.004 \nonumber \\
                  (sin 2\theta_{2s})^2 &=& 0.2 \; ,
\eeq
and the masses are negligibly small. Note that this is in contrast to the
parameters of Ref.\cite{fkmp03}, with the constraint of dark matter giving
a mixing angle of $(sin 2\theta_{dm})^2 \simeq 10^{-8}$, and a mass greater
than 1 keV.  In our present work we will use values for $(sin 2\theta)^2$ 
in the range 0.2 to 0.004 to estimate the pulsar kick, which is compatible 
with the recent global analysis\cite{kdcss09}

The probability of asymmetric emission, giving a pulsar kick,
does not depend directly on the sterile neutrino mass in our model, but
is proportional to the $(sin 2\theta_s)^2$. It is the large mixing angles
found in fits to MiniBoone and LSND that lead us to carry out the
investigation in the present paper. 

\subsection{General Formulation of Neutrino Emissivity With a Strong Magetic
Field}
   The neutrino emissivity is given in general form in many 
papers, e.g., see Refs~\cite{fm,bw}:
\beq
\label{6}
   e^\nu &=&\Pi^4_{i=1} \int \frac{d^3p^i}{(2\pi)^3}
\frac{d^3 q^\nu}{2 \omega^\nu(2 \pi)^3} 
 \int\frac{d^3q^e}{(2\pi)^3}
\nonumber \\
 && (2\pi)^4 \sum_{s_i,s^\nu} \frac{1} {2\omega^e_L}  \omega^\nu \mathcal{F}
M^{\dagger} M \\
&&  \delta(E_{final}-E_{initial}) \delta(\vec p_{final}-
\vec p_{initial})  \nonumber \; ,
\eeq
where $M$ is the matrix element for the URCA process and
 $\mathcal{F}$ is the product of the initial and final Fermi-Dirac 
functions corresponding to the temperature and density of the medium.
The main source of the asymetric emissivity that produces the pulsar
velocity is the fact that the electron has a large probability to be
in the lowest (n=0) Landau level. See Refs~\cite{jl,mo} for a discussion 
of Landau levels. The asymetric emissivity can be seen by considering the
weak axial intereaction, $W_A$,
\beq
\label{7}
   W_A &=& -\frac{G}{\sqrt{2}}g_A\chi_p^\dagger \vec{l}\cdot \vec{\sigma}
\chi_n \\
           l_\mu &=& \bar{\Psi}(q^e) \gamma_\mu(1-\gamma_5)\Psi(q^\nu) 
\nonumber \; ,
\eeq
with $G=\frac{10^{-5}}{m_n^2},\;  g_A=1.26$, the $\chi$  are
the nucleon spinors, and the lepton wave functions are $\Psi(q^e),\Psi(q^\nu)$,
where  $q^e$ and $q^\nu$ are the electron and antineutrino momenta, 
respectively. The key to the asymmetric emission is given be the trace
over the leptonic currents, $Tr[l_i^{\dagger} l_j]$,
\beq
\label{8}
 \int d^2 q^e_\perp Tr[l_i^{\dagger} l_j] & \simeq & 8 \pi E^e [(q^\nu)^j 
\delta_{i3} +(q^\nu)^i \delta_{j3} \nonumber \\
  &&-\delta_{ij} (q^\nu)^3]  (\hat{q}^e = \hat{B} = \hat{z}) \; ,
\eeq
with the magnetic field B in the z direction. We only consider the weak
axial force, which is dominant. Using the relationship given in Eq(\ref{8})
one can show that the result of the traces and integrals over the axial 
product matrix element has the form ($\hat{B} = \hat{z}$)
\beq
\label{9}
   \int\int |M_A|^2 &\propto& (q^\nu)^z \; .
\eeq
Details are given in Ref~\cite{hjk07}, where it is shown that the asymmetric
neutrino emissivity, using the general formulation of Ref~\cite{fm}, is
\beq
\label{10}
  \epsilon^{AS} &\simeq& 0.64 \times 10^{21} T_9^7 P(0)\times f
\nonumber \\ 
&&{\rm erg\; cm^{-3}\; s^{-1}}=p_{ns} c \frac{1}{V_{eff} \Delta t} \; ,
\eeq
where $T_9 = T/(10^9 K)$, $p_{ns}$ is the neutron star momentum, $P(0)$ is 
the probability of the electron produced with the antineutrino
being in the lowest Landau state, f=.52 is the probability of the
neutrino being at the + z neutrinosphere surface~\cite{hjk07}, $V_{eff}$ is
the volume at the surface of the neutrinosphere from which neutrinos are
emitted, and $\Delta t \simeq 10 s$ is the time interval for the emission.
 
We derive $P(0)$ and $V_{eff}$, the effective volume for the emissivity, 
in the next two subsections.

\subsection{$P(0)$ = Probability for the Electron to be in the n=0 Landau 
Level}

Just as in our previous work in which Landau levels play a crucial 
role\cite{hjk07}, only the lowest Landau level, for which the helicity
is -1/2 (rather than $\pm$1/2 as with the usual Dirac spinors) gives
asymmetric emission. The probability that an electron in a strong magnetic
field is in the lowest (n=0) Landau level, $P(0)$, can be calculated from the
temperature, T, and the energy spectrum of Landau levels\cite{jl,mo}.
A particle with momentum p and effective mass $m_e^*$ in a magnetic field 
B in the nth Landau level has the energy

\beq
\label{11}
           E^L(p,n) &=&  \sqrt{p^2 +(m_e^*)^2 + 2 (m_e^*)^2 \frac{B}{B_c} n} 
\; ,
\eeq
with $B_c = 4\times 10^{13}$ G, and $m_e^*$ is the 
effective mass of the electron at the high density of the protoneutron 
star and neutrinosphere.  
 
 From standard thermodynamics the probability of occupation of the n=0
Landau level, P(0), is given by\cite{fkmp03}:
\beq
\label{12}
   P(0) &=& \frac{F(0)}{F(0) + 2 \sum_{1}^{\infty} F(n)} \; ,
\eeq
where F(n), with magnetic field B, temperature T, and chemical potential $\mu$,
is 
\beq
\label{13}
       F(n) &=& \int_{p_{min}}^{\infty} dp \frac{[m_n-m_p-E^L(p,n)]^2}
{1+exp{[(E^L(p,n)-\mu)/T]}} \; .
\eeq
The electron energy is restricted to magnitudes greater than $\mu$,
but the integrals in Eq.(\ref{7}) are insensitive to
$p_{min}$, so we take $p_{min}$=0 as in Ref\cite{fkmp03}.

 We agree with the estimate of Ref.\cite{fkmp03} for P(0). Note that if we
had used the free electron mass, $m_e$, in the Landau energies (Eq.(\ref{5}))
we would have obtained a much smaller value for P(0).
For B= $10^{16}$ G, $\mu=40$ MeV, $m_e^*$=4 MeV
and $T_{\nu-sphere}$ = 20 MeV, P(0) $\simeq$ 0.3. This is similar to our
estimate of P(n=0) $\simeq$ 0.4 at the surface of the protoneutron star
at about 10 seconds\cite{hjk07}. Therefore our result for asymmetric
emissivity differs from that of Fuller et al \cite{fkmp03} mainly in that 
we have a much larger mixing angle, and a much smaller effective volume, since
the sterile neutrinos oscillate back to active neutrinos within the
neutrinosphere; and therefore our emission only takes place near the
surface of the neutrinoshere. However, in contrast with purely active
neutrino emission in which the opacity results in very small pulsar
kicks~\cite{lq98}, the sterile neutrinos have a much larger effective
volume, and can therefore produce much larger pulsar velocities

\subsection{Estimate of $V_{eff}$= Effective Volume for Emission}

  To estimate $V_{eff}$ we make use of the early study of opacity in about
the first 20s of the creation of a neutron star via a supernova 
collapse~\cite{ip82,bl86,bml81}, and a recent detailed study of neutrino 
mean free paths\cite{ss07}. Since the mean free path of the sterile neutrino is
determined by that of the standard neutrino to which it oscillates, $\lambda$,
we make use of studies of active neutrino mean free paths. First note
that the neutrino mean free path is given by 
\beq
\label{14}
          1/\lambda &=& \int \frac{d^3 p}{(2 \pi)^3} M_{fi} [1-n(q)]
\nonumber \\
        &&  \times (1+ e^{(\mu_\nu -E_\nu)/kT}) \;,
\eeq
where $M_{fi}$ is the weak matrix element and $n(q)$ is the Fermi distribution.
For the calculation of the sterile neutrino, for which $M_{fi}=0$,
one can use the value of $1/\lambda$ with a factor of $sin^2(2\theta)$ from
the matrix element and another such factor from the occupation probability.
From the results of previous authors, for T in the 10 to 20 Mev range and 
$\mu$ in the 20 to 40MeV range, we estimate that  $\lambda \simeq 1.0 cm$ 
This gives a range for the effective sterile neutrino mean free path
\beq
\label{15}
        \lambda_s &\simeq& 5.0 {\rm \; to\;} 250 {\rm \; cm}  \; .
\eeq 

For a neutrinosphere radius of 40 km, with $\lambda_s << R_\nu$ this gives us 
$V_{eff}= (4\pi/3)( R_\nu^3-(R_\nu-\lambda_s)^3) \simeq 4\pi R_\nu^2 
\lambda_s$.

  From Eq.({\ref{10}), $R_\nu$ = 40 km, and $\lambda$=1.0 cm, 
\beq
\label{16}
   p_{ns}&=& M_{ns} v_{ns} \simeq \frac{0.67 \times 10^{25}}{ sin^2(2 \theta)}
 T_9^7 gm \frac{cm}{s} \; ,
\eeq
with $T_9 = \frac{T}{10^9 K}$.
Taking the mass of the neutron star to equal the mass of our sun, $M_{ns}
= 2 \times 10^{33}$ gm, we obtain for the velocity of the neutron star
\beq
\label{17}
        v_{ns} &\simeq& 3.35 \times 10^{-7} (\frac{T}{10^{10} K})^7 
\frac{1}{sin^2(2\theta)} \frac{km}{s} \; ,
\eeq.

For example, for T=10 Mev =$1.16\times 10^{11}$ kK,
\beq
\label{18}
             v_{ns} &\simeq& 47.3 \frac{km}{s} {\rm \; to \; }
2,370 \frac{km}{s} \; ,
\eeq
which means that sterile neutrino emission could account for the large
pulsar kick with the parameters extracted from Refs~\cite{ms07,s07}. If we
use the physical parameters that give Eq.(\ref{18}) for electron neutrinos,
we obtain a pulsar velocity of $v_{ns}$ = 95 km/s, which is consistent
with previous predictions by several authors.

  It should be noted that the study of the MiniBooNE and LSND results are
in progress, and the mixing angles that result could be different from
those obtained in Refs.\cite{ms07,s07}. For this reason we have used a
range of parameters. Preliminary data from the MiniBooNE/Minos 
experiment\cite{minos08}, however, is consistent with the MiniBooNE 
results\cite{mini}. We also once more point out that although Ref~\cite{mm08} 
questions the accuracy of the peramaters extracted by Ref~\cite{ms07}, our 
model is compatible with the recent global analysis\cite{kdcss09}, with 
very good $\chi^2$ probabilities.

\section{Conclusions}

   Because of the strong magnetic fields in protoneutron stars and the
associated neutrinosphere, the electrons produced in the URCA processes
that dominate neutrino production in the first 10 seconds have a sizable 
probability, P(0), to be in the lowest (n=0) Landau level. This leads to 
asymmetric neutrino momentum. With the mixing angles found in 
Refs\cite{ms07,s07}, we find that the sterile neutrinos produced during 
this period for high luminosity pulsars can give the pulsars velocities of 
greater than 1000 km/s, as observed. We emphasize that our results are
consistent sterile neutrino parameters based on a global analysis of
data from MiniBoonE, LSND, KARMEN, NOMAN, Bugey, CHOOZE, CCFR84, and 
CDHS\cite{kdcss09}, rather than a model. There is a high probability 
that light sterile neutrinos with a large mixing angle exist,
which is the basis of our work.

   There is a strong correlation of the pulsar velocity with temperature, T.
Since it is difficult to determine T accurately, it is difficult for us to
predict the velocity of a pulsar whose kick arises from sterile neutrino
emission. On the other hand, if the pulsar kick arises from the asymmetric
emission of active neutrinos produced by the modified URCA processes after
10 seconds, also proportional to P(0)\cite{hjk07}, then T can be determined
by an accurate measurement of the neutrinos from the supernova. Therefore,
in future years, with much more accurate neutrino detectors, one could predict 
the velocity of the resulting pulsar. Unfortunately, the energy of emitted 
sterile neutrinos cannot be measured. From our results in the present paper
and those in Ref\cite{hjk07}, high luminosity pulsars receive a large kick 
both from sterile neutrinos in the first ten seconds and standard neutrinos 
in the second ten seconds.
\newpage
pulskicksterile8-5-09
\hspace{2cm} {\bf Acknowledgements}
\vspace{5mm}

This work was supported in part by DOE contracts W-7405-ENG-36 and 
DE-FG02-97ER41014. The authors thank Terry Goldman; and William Louis, 
Gerald Garvey and other LANL members of the MiniBooNE Collaboration for 
helpful discussions.

\end{document}